\documentclass[12pt,preprint]{emulateapj}
\usepackage{graphicx}

\shortauthors{Dragomir et al.}
\begin{document}

\renewcommand{\bottomfraction}{0.9}

\title{Rayleigh Scattering in the atmosphere of the warm exo-Neptune GJ 3470b}

\author{
  Diana Dragomir\altaffilmark{1,2,3},
  Bj{\"o}rn Benneke\altaffilmark{4},
  Kyle A. Pearson\altaffilmark{5},
  Ian J. M. Crossfield\altaffilmark{6,7},
  Jason Eastman\altaffilmark{8},
  Travis Barman\altaffilmark{6},
  Lauren I. Biddle\altaffilmark{9}
  }

\email{diana@oddjob.uchicago.edu}
\altaffiltext{1}{Las Cumbres Observatory Global Telescope Network, 6740 Cortona Dr. Suite 102, Goleta, CA 93117, USA}
\altaffiltext{2}{Department of Physics, Broida Hall, UC Santa Barbara, CA, USA}
\altaffiltext{3}{The Department of Astronomy and Astrophysics, University of Chicago, 5640 S Ellis Ave, Chicago, IL 60637, USA}
\altaffiltext{4}{Division of Geological and Planetary Sciences, California Institute of Technology, Pasadena, CA 91125, USA}
\altaffiltext{5}{Department of Physics and Astronomy, Northern Arizona University, Flagstaff, AZ 86001}
\altaffiltext{6}{Department of Planetary Sciences, Lunar and  Planetary Laboratory, University of Arizona, Tucson, AZ, 85721, USA}
\altaffiltext{7}{Sagan Fellow}
\altaffiltext{8}{Harvard-Smithsonian Center for Astrophysics, Cambridge, MA 02138 USA}
\altaffiltext{9}{Gemini Observatory, Northern Operations Center, 670 N. Aohoku Place, Hilo, HI 96720, USA}

%%%%%%%%%%%%%%%%%%%%%%%%%%%%%%%%%%%%%%%%%%%%%%%%%%%%%%%%%%%%%%%%%%%%

\begin{abstract}

GJ 3470b is a warm Neptune-size planet transiting a M dwarf star. Like the handful of other small exoplanets for which transmission spectroscopy has been obtained, GJ 3470b exhibits a flat spectrum in the near- and mid-infrared. Recently, a tentative detection of Rayleigh scattering in its atmosphere has been reported. This signal manifests itself as an observed increase of the planetary radius as a function of decreasing wavelength in the visible. We set out to verify this detection and observed several transits of this planet with the LCOGT network and the Kuiper telescope in four different bands (Sloan g', Sloan i', Harris B and Harris V). Our analysis reveals a strong Rayleigh scattering slope, thus confirming previous results. This makes GJ 3470b the smallest known exoplanet with a detection of Rayleigh scattering. We find that the most plausible scenario is a hydrogen/helium-dominated atmosphere covered by clouds which obscure absorption features in the infrared and hazes which give rise to scattering in the visible. Our results demonstrate the feasibility of exoplanet atmospheric characterization from the ground, even with meter-class telescopes.

\end{abstract}

\keywords{planetary systems -- techniques: photometric -- stars: individual (GJ~3470)}

%%%%%%%%%%%%%%%%%%%%%%%%%%%%%%%%%%%%%%%%%%%%%%%%%%%%%%%%%%%%%%%%%%%%

\section{Introduction}
\label{introduction}

In the two decades since the first detection of an extrasolar planet orbiting a main sequence star \citep{May95}, exoplanets have presented an uninterrupted stream of unexpected surprises. From isolated hot Jupiters on retrograde orbits to tightly-packed, multi-planet systems of super-Earths, none seem to quite resemble the Earth nor to be arranged in a structure similar to that of our Solar System. Hundreds of new transiting planets continue to be discovered each year in data from surveys such as Kepler \citep{Bor10}, K2 \citep{HS14}, SuperWASP \citep{Pol06} and HATNET/HATSouth \citep{Bak04, Bak13}, to name only a few. While for hot Jupiters it is possible to measure the planetary mass using radial velocity measurements, this is more challenging for smaller planets, though still possible for those in small orbits around bright stars. Even so, the density of planets with radii approximately between that of Earth and Neptune does not uniquely determine their composition. Detailed atmospheric characterization of the most observationally accessible of these systems is necessary to gain deeper insight into the nature of these objects. In principle, such observations (transmission and emission spectroscopy at a wide range of wavelengths) can distinguish between different planetary compositions -- rocky, water-rich, or hydrogen-dominated. These studies will continue to emphasize the uniqueness of our own Solar System by constraining the composition, structure, and formation of extrasolar systems.

Though Jovian-sized planets were the first discovered by both radial velocity and transit observations, smaller exoplanets are of especial interest because a wide range of compositions, particularly atmospheric composition, is possible for planets with masses comparable to and smaller than that of Neptune. Understanding the nature of increasingly smaller exoplanets also constitutes a stepping stone toward determining how frequent truly Earth-like planets are, and how they form. The Kepler mission has discovered over 4000 planet candidates \citep{Mull15}, the vast majority of which are similar in size to or smaller than Neptune. After accounting for detection biases, planets in this size category occur $\sim 10\times$ more frequently per $\log R_P$ than larger planets, and are 2--3$\times$ more common around M dwarfs than around Sunlike stars \citep{How12, Mul15}.  

Despite the large numbers of known or candidate planets, and despite the high frequency of small planets around main sequence stars, the host stars of most systems are too faint for the atmospheric makeups of their planets to be studied. A host star that is very bright will facilitate such studies, even for transit depths of the order of 0.1$\%$. Notable examples are 55 Cnc e (\citealt{Win11, Dem11}; V = 5.95) and HD 97658b (\citealt{Dra13}; V = 7.7). Small planets transiting somewhat fainter stars are also accessible to existing facilities if their transits are deeper than a couple of 0.1$\%$, which could occur due to a smaller host star and/or a larger planet. GJ 1214b (\citealt{Cha09}; V = 14.7), GJ 436b (\citealt{Gil07}; V = 10.6) and GJ 3470b (\citealt{Bon12}; V = 12.3), all of which orbit M dwarf (so smaller) stars, are such systems. This small, but growing, sample of favorable super-Earths, sub-Neptunes, and hot Neptunes continues to be the target of considerable effort to probe these planets' atmospheres via transmission spectroscopy \citep{Sea00,Bro01}. So far, these observations often reveal featureless transmission spectra, which are interpreted as a high-mean-molecular-weight atmosphere and/or high-altitude clouds or hazes \citep{Mil09,Kre14}.

A few groups \citep{Pon08, Sin11, Sin15, Nar13, Moo13} have attempted to break this degeneracy by seeking the Rayleigh-scattering slope expected to dominate both hazy and clear atmospheres at the shortest optical wavelengths. With an estimated planetary temperature and gravity, the detection of a Rayleigh scattering slope yields the atmospheric scale height and thus empirically determines the atmosphere's mean molecular weight. Until now, this measurement has only been attempted for a few planets. This measurement can also be useful for hot Jupiters since some also show indications of high altitude clouds, and indeed these studies have mostly been performed for giant exoplanets because a Rayleigh scattering signal would be easier to detect for planets with larger scale heights. An increase in planetary radius with decreasing wavelength in the atmosphere of the hot Jupiter HD 189733b is reported by \cite{Pon08} and \cite{Sin11}, based on HST ACS and STIS observations, respectively. After taking into account the overall flux level of the star and implementing a wavelength-dependent radius correction assuming a reasonable spot temperature, \cite{Pon08} and \cite{Sin11} demonstrate the effect is planetary rather than stellar, and they attribute it to the presence of Rayleigh scattering due to high-altitude hazes. While \cite{McC14} find that such a rise in planetary radius could, in the case of HD 189733, be explained by latitudinal bands of star spots, \cite{Pon13} have demonstrated that the stellar variability of HD 189733 is caused by isolated spots moving in and out of view. These studies emphasize the importance of adequately characterizing the stellar variability and the spot properties when undertaking such analyses. Recently, a detection of Rayleigh scattering was also announced for other hot Jupiters, including WASP-6b \citep{Jor13, Nik15} and WASP-31b \citep{Sin15}. \cite{Nar13} and \cite{Moo13} have carried out such observations for GJ 1214b, a warm super-Earth orbiting a M dwarf, but found the transmission spectrum to be as featureless at these short wavelengths as it is at longer wavelengths \citep{Ber12,Fra13}.

%%%%%%%%%%%%%%%%%%%%%%%%%%%%
%%%%%%%%Observing log table%%%%%%%%%%%

\begin{deluxetable*}{lccccc}[!h]
\tabletypesize{\footnotesize}
\tablecaption{Observing Log}
\tablewidth{18cm}
\tablehead{\colhead{Observing Date} & \colhead{Telescope size/site} & \colhead{Filter} & \colhead{Exposure time} & \colhead{Airmass} & \colhead{Standard deviation} \\
\colhead{(UT)} & \colhead{} & \colhead{} & \colhead{(s)} & \colhead{} & \colhead{of residuals (mmag)}} %\\
\startdata

2013 March 19 & 1.0m (LCOGT - McDonald) & Sloan i' & 45 & 1.08 $\rightarrow$ 1.04 $\rightarrow$ 2.35 & 2.1  \\
2013 December 30 & 1.0m (LCOGT - McDonald) & Sloan i'  & 45 & 1.98 $\rightarrow$ 1.04 & 1.6 \\
2014 January 3 & 2.0m (LCOGT - Siding Spring)  & Sloan g' & 240 & 2.34 $\rightarrow$ 1.46 $\rightarrow$ 1.48 & 1.4 \\
2014 January 19 & 1.5m (Kuiper) & Harris V & 40 & 1.70 $\rightarrow$ 1.05 & 1.8  \\
2014 January 22 & 2.0m (LCOGT - Haleakala) & Sloan g' & 240 & 1.01 $\rightarrow$ 2.17  & 3.4 \\ 
2014 January 26 & 1.0m (LCOGT - SAAO) & Sloan i'  & 45 & 1.84 $\rightarrow$ 1.49 $\rightarrow$ 1.86 & 2.7 \\
2014 February 2 & 2.0m (LCOGT - Siding Spring) & Sloan g' & 240 & 1.64 $\rightarrow$ 1.46 $\rightarrow$ 2.06  &  0.7 \\
2014 February 8 & 1.5 (Kuiper) & Harris B & 108 & 1.23 $\rightarrow$ 1.05 $\rightarrow$ 1.15 & 1.8 \\
2014 February 19 & 1.0m (LCOGT - McDonald) & Sloan i'  & 45 & 1.06 $\rightarrow$ 1.04 $\rightarrow$ 1.45 &  1.9 \\
2014 February 25 (a) & 1.0m (LCOGT - SAAO) & Sloan i'  & 45 & 1.58 $\rightarrow$ 1.49 $\rightarrow$ 1.95 & 2.8 \\
2014 February 25 (b) & 1.0m (LCOGT - SAOO) & Sloan i'  & 45 & 1.58 $\rightarrow$ 1.49 $\rightarrow$ 1.93 & 2.9 \\
2014 March 1 & 1.0 (LCOGT - McDonald) & Sloan g'  & 45 & 1.22 $\rightarrow$ 1.04 $\rightarrow$ 2.83 & 2.9  \\

\enddata
%\tablenotetext{}{}
\label{tab:obs_log}
\end{deluxetable*}

%%%%%%%%Observing log table%%%%%%%%%%%
%%%%%%%%%%%%%%%%%%%%%%%%%%%%

The first sub-Jovian planet with any significant features reported in transmission is the subject of our study.  GJ~3470b is a `warm Neptune' with $T_{eq} \approx$ 700 K, $R_P=3.9 R_{\oplus}$ and $M_P=13.7 M_{\oplus}$ \citep{Bid14} transiting exoplanet in orbit around a M1.5V star. Its low density of 1.18 $\pm$ 0.18 g cm$^{-3}$ strongly points to a substantial atmosphere covering the planet. Recent Large Binocular Telescope (LBT) observations of a single transit of GJ 3470b in dual-band photometry indicates the presence of a strong Rayleigh scattering slope \citep{Nas13} despite the absence of strong absorption in the infrared \citep{Cro13, Ehr14}. The optical transit photometry was obtained simultaneously through the LBT's {\it U$_{spec}$} and {\it F97N20} filters, centred at 357.5 and 963.5 nm, respectively. A 6$\sigma$ difference was observed between the transit depths in the two filters, indicating a significant detection of Rayleigh scattering. It is not clear whether the atmosphere of GJ3470b is highly metal-enhanced, nearly pure H$_2$, or somewhere in-between. On the one hand, \cite{For13} find that, based on population synthesis models, low-mass low-density planets such as GJ 3470b are likely to form with a high metallicity H/He atmosphere. This conclusion is supported by other planet formation studies as well \citep{Mor12, Fig09}. On the other hand, the LBT observations tentatively indicate a haze-covered but low-metallicity H/He-dominated atmosphere.

If confirmed, the results of \cite{Nas13} suggest that it may indeed be possible to characterize the atmospheres of exoplanets using shorter-wavelength measurements even when their IR transmission spectra are featureless, an exciting prospect. However, their conclusions are based on photometry obtained during a single transit, and their photometric uncertainties do not take into account the presence of correlated noise in the light curves. In addition, there is a need for transit measurements at wavelengths between the two wavelengths probed by \cite{Nas13} in order to better constrain the prospective Rayleigh scattering feature in the 400 - 900 nm region. These factors motivate additional observations in this wavelength regime. In this work, we present new, four-color, broadband photometry acquired during several transits of GJ~3470b with which we aim to improve constraints on this planet's short-wavelength Rayleigh scattering slope and its atmospheric composition. We describe the observations in Section \ref{sec:Phot} and the light curve analysis in Section \ref{sec:Analysis}. Our modelling procedure and results are presented in Section \ref{sec:Res}. We discuss the implications of our results in Section \ref{sec:Disc} and conclude in Section \ref{sec:Conc}.

%%%%%%%%%%%%%%%%%%%%%%%%%%%%%%%%%%%%%%%%%%%%%%%%%%%%%%%%%%%%%%%%%%%%

\section{Photometric observations}
\label{sec:Phot}

We acquired new observations during eleven transits of GJ 3470b, between December 2013 and March 2014. These consist of seven full and four partial transits, for all of which we have light curves covering more than half a transit (and including at least ingress or egress). We describe below the observations and data reduction for all of these transits, including four obtained in the Sloan g', five in the Sloan i', one in the Harris B and one in the Harris V filters. We also include in our analysis a sixth Sloan i' transit that was first published in \cite{Bid14}. A detailed observation log is found in Table \ref{tab:obs_log}.

\subsection{Kuiper observations}

We obtained two transit light curves with the Steward Observatory 1.55m Kuiper Telescope on Mt. Bigelow, using the Mont4k CCD. The Mont4k CCD contains a 4096 x 4096 pixel sensor with a field of view of 9.7'x9.7'. One of the transit light curves was observed through the Harris-B (350-550nm) filter and the other through the Harris-V (475-675 nm) filter. The exposure times used were 108 and 40 seconds, respectively. The B and V band measurements were obtained on 2014 February 8 and 2014 January 14, respectively. 

Each of the Kuiper datasets were reduced using the Exoplanet Data Reduction Pipeline, described in \cite{Pea14}. The pipeline generates a series of IRAF scripts that will calibrate images using the standard reduction procedure and perform aperture photometry. To produce the light curve we performed aperture photometry (using the task PHOT in the IRAF DAOPHOT package) by measuring the flux from our target star as well as the flux from various reference stars at different aperture radii. The aperture radii and comparison star that produced the lowest scatter in the out of transit data points were used to produce the final GJ 3470b light curves. Because GJ 3470 is very bright and the Kuiper Telescope field of view is relatively small, one reference star was ultimately used for the differential photometry.

\subsection{LCOGT observations}

Ten of our transit light curves (including one already published in \cite{Bid14}) were obtained using the LCOGT network, which currently consists of nine 1.0m and two 2.0m telescopes (Faulkes Telescope North and Faulkes Telescope South) located at five longitudinally distributed sites in both the southern and northern hemispheres. Thanks to its low declination (RA = 07 59 05.87; Dec = +15 23 29.5), it was possible to observe GJ 3470 from either hemisphere. All of the i' band and one of the g' band transits were observed with 1.0m telescopes, while the remaining three g' band transits were observed with 2.0m telescopes. The 1.0m and 2.0m data were obtained using SBIG STX-16803 4096 x 4096 cameras with a 15.8' x 15.8' FOV, and Fairchild CCD486 BI 4096 x 4096 spectral imaging cameras with a 10.5' x 10.5' FOV, respectively.

Observations taken in the i' band had an exposure time of 45 s and were obtained with the telescope defocussed in order to avoid saturation and to increase the open shutter time relative to the overhead time (thus improving the duty cycle). The defocus was 2.8mm from the focal plane. The g' band observations were taken with exposure times of 240 s and 45 s on the 2.0m and 1.0m telescopes, respectively. While it may seem counterintuitive to expose longer when employing a larger telescope, we note that on the nights during which we observed with the 2.0m's, the local seeing was greater than 2''. In addition, on the night of the 1.0m g' transit (observed at McDonald Observatory), the seeing was less than 2'' and the transparency was exceptionally high.

All images were processed using the pipeline described in \cite{Bro13}. The light curves were extracted using pyraf aperture photometry routines. Differential photometry was carried out using the same ensemble of comparison stars for a given filter. Two and three reference stars were used to calibrate the g' band and i' band photometry of GJ 3470, respectively. 

%%%%%%%%%%%%%%%%%%%%%%%%%%%%
%%%%%%%%BIC table%%%%%%%%%%%

\begin{deluxetable*}{lcccccc}[!h]
\tabletypesize{\footnotesize}
\tablecaption{Bayesian Information Criterion Table}
\tablewidth{13cm}
\tablehead{\colhead{Transit date} & \colhead{No detrending} & \colhead{am} & \colhead{t} & \colhead{p} & \colhead {p + am} & \colhead{p + t}}
\startdata

2013 March 19  & \textbf{760} & 759 & 779 & - & - & - \\ 
2013 December 30 & 567 & \textbf{550}  & 1197 & - & - & - \\
2014 January 3 & 1374 & - & - & 619 & 851 & \textbf{536} \\
2014 January 19 & \textbf{771}  & 772 & 809 & - & -  \\
2014 January 22 & \textbf{248} & 247 & 252 & - & - & - \\
2014 January 26 & 1740 & 1735  & \textbf{1730} & - & - & -  \\
2014 February 2 & 2415 & \textbf{166} & 1384 & -  & - & -  \\
2014 February 8 & 233 & 223 & \textbf{179} & - & - & -  \\
2014 February 19 & \textbf{559} & 562 & 616 & - & - & - \\
2014 February 25 (a) & 1607 & \textbf{1367} & 1450 & - & - & - \\
2014 February 25 (b) & 1573 & 1406 & \textbf{1395} & - & - & - \\
2014 March 1 & 792 & - & - & 876 & \textbf{727} & 809 \\

\enddata
\tablenotetext{-}{Bayesian Information Criterion (BIC) values for no detrending and different combinations of de-trending parameters. We only de-trended for x or y position only if there was an obvious drift ($>$ 1 pixel) in the star positions through a given transit. Column header symbols are as follows: $am$ is a linear function of the airmass, $p$ is the x and y pixel position and $t$ is a time-dependent linear trend. The de-trending models selected for each transit are marked in bold.}
\label{tab:BIC}
\end{deluxetable*}

%%%%%%%%BIC table%%%%%%%%%%%
%%%%%%%%%%%%%%%%%%%%%%%%%%%%

%%%%%%%%%%%%%%%%%%%%%%%%%%%%%%%%%%%%%%%%%%%%%%%%%%%%%%%%%%%%%%%%%%%%

\section{Light curve analysis}
\label{sec:Analysis}

Most of the transit light curves required de-trending from one or more of the following parameters: time, airmass, x and y position on the detector. To avoid overfitting, we de-trended for x or y position only if there was an obvious drift ($>$ 1 pixel) in the star positions through a given transit. %We tested both linear and quadratic functions of time. 

The transits were analysed with EXOFAST \citep{Eas13}, a differential evolution Markov Chain Monte Carlo (MCMC) algorithm. Each transit was fit (using the \cite{Man02} transit model) simultaneously with the de-trending parameters. We used the Bayesian Information Criterion (BIC) for model selection, and selected the combination of de-trending parameters (if any) that resulted in the lowest BIC value by at least $\Delta BIC = 2$ (indicating positive evidence for that particular combination; \citealt{Kas95}). Table \ref{tab:BIC} shows the BIC values for the combinations of de-trending parameters we tested. EXOFAST scales photometric uncertainties using the method of \cite{Gil10} which multiplies the photometric uncertainties by the ratio of the RMS of the binned light curve to the RMS expected if only white noise was present in the data. This scaling factor is determined based on the light curve residuals.

\begin{figure*}[!h]
\begin{center}
\includegraphics[scale=0.28]{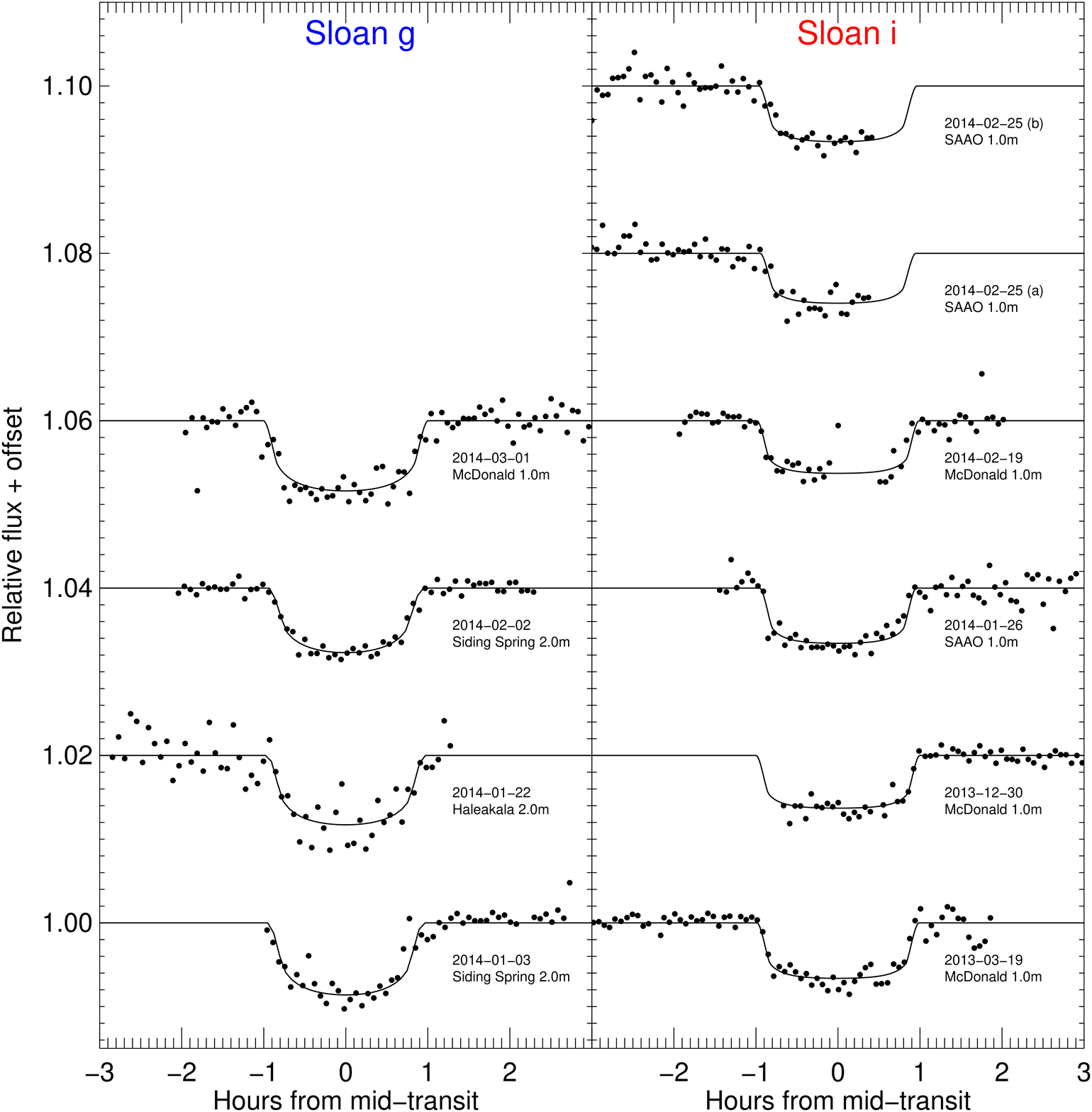}
\caption{Four transits observed in Sloan g' band ({\it left}) and six transits observed in Sloan i' band ({\it right}) with the LCOGT network, binned in 4-minute intervals. All observations, with the exception of those from March 19, 2013, were obtained over a period of just over three months. Best-fitting models for each individual transit are shown. The associated best-fit parameters are listed in Table \ref{tab:GJ3470b_LCOGT}.}
\label{fig:gpip}
\end{center}
\end{figure*}

\begin{figure}[!t]
\begin{center}
\includegraphics[scale=0.34]{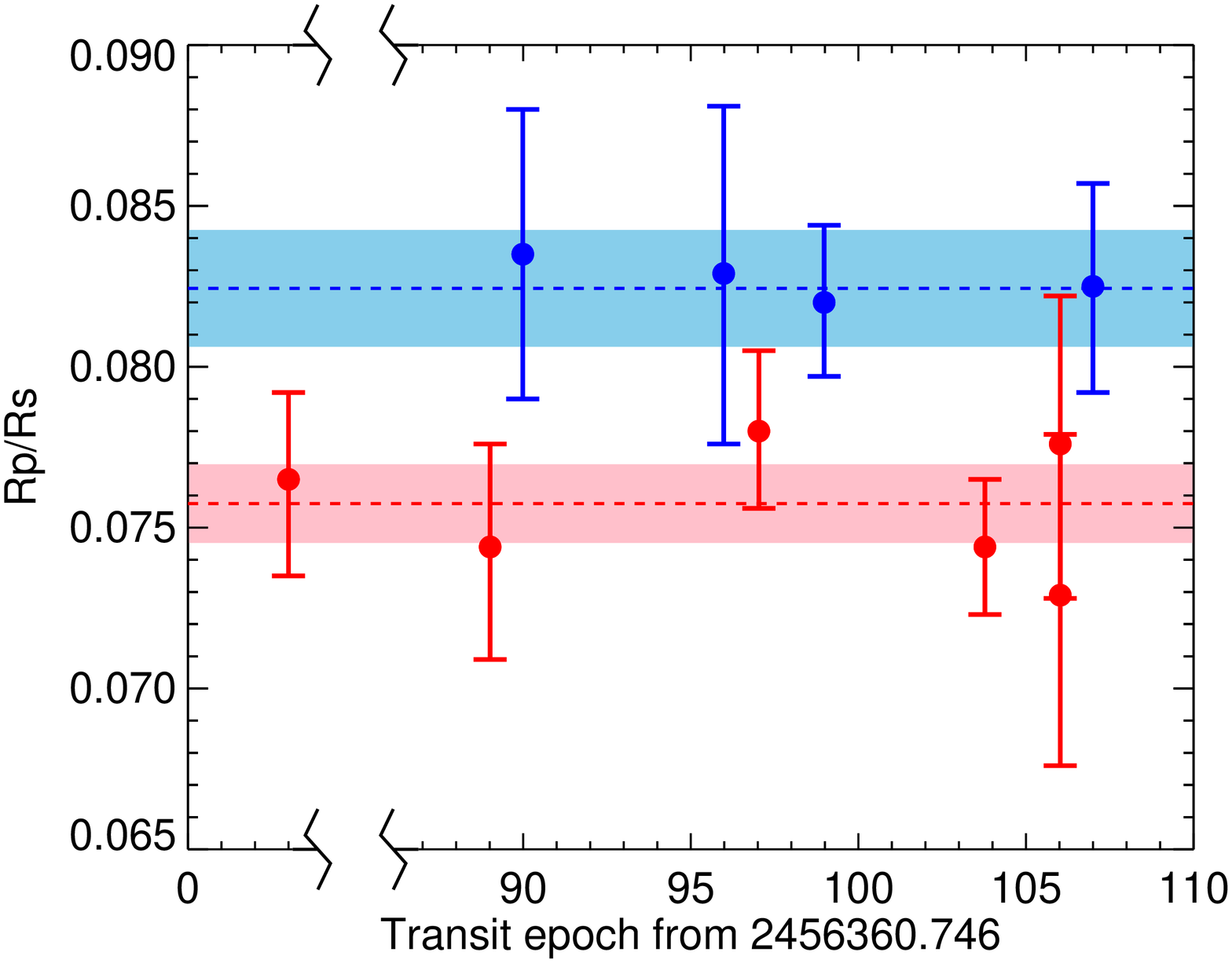}
\caption{$R_p/R_s$ as a function of transit epoch, for each of the LCOGT transits analyzed in this work. Values for Sloan g' transits are shown in blue and for Sloan i' transits in red. The dashed lines and colored bands correspond to the weighted mean of the $R_p/R_s$ values and the uncertainty in the mean, respectively, in each filter. A colour version is available in the online version of the journal.}
\label{fig:gpipcomp}
\end{center}
\end{figure}

We allowed the mid-transit time ($T_{0}$), period ($P$), scaled semi-major axis ($a/R_{S}$), orbital inclination ($i$) and the planet-to-star radius ratio ($R_{P}/R_{S}$) to vary as free parameters. For each of the four filters used, we computed quadratic limb darkening coefficients using the methods described in \cite{Cro13} and the stellar parameter values of \cite{Dem13}. The calculations were based on a spherically symmetric PHOENIX stellar atmosphere model. We estimate uncertainties for the limb darkening values by varying the stellar parameters within their own uncertainties. Our data are not sufficiently precise to fit for the limb darkening coefficients, so the transit analysis is performed with the coefficients fixed to the values we calculated. These calculated limb darkening coefficients and the fitted transit parameter values for each transit are found in Tables \ref{tab:GJ3470b_LCOGT} and \ref{tab:GJ3470b_Kuiper}. We note that all of our values of $P$, $a/R_{S}$ and $i$ agree with those of \cite{Dem13} to within 1$\sigma$, and with those of \cite{Bid14} to within 1 to 2$\sigma$. The individual g' and i' transits are shown in Figure \ref{fig:gpip}, and the best-fit values of $R_{P}/R_{S}$ for each transit are plotted in Figure \ref{fig:gpipcomp}.

%%%%%%%%%%%%%%%%%%%%%%%%%%%%
%%%%%%%%%%Parameter tables%%%%%%%%%%

%%%%%%%%%%LCOGT Table%%%%%%%%%%

\newcommand{\bjdtdb}{\ensuremath{\rm {BJD_{TDB}}}}

\begin{deluxetable}{lcc}
\tabletypesize{\footnotesize}
\tablecaption{Best-fit parameters found for GJ 3470\lowercase{b} LCOGT transits}
\tablehead{\colhead{~~~Parameter} & \colhead{Value (g')} & \colhead{Value (i')}}
\startdata
    ~~~ $\lambda$ (nm) & 490 & 763  \\
	~~~ ${\it u_1}$ & ${\it 0.398}\pm{\it 0.050}$ & ${\it 0.123}\pm{\it 0.050}$ \\
           ~~~ ${\it u_2}$ & ${\it 0.390}\pm{\it 0.050}$ & ${\it 0.489}\pm{\it 0.050}$ \\

\sidehead{Transit 6:}
       ~~~ $T_C$ (\bjdtdb) & - & $6714.4284_{-0.0048}^{+0.0046}$ \\
~~~ $R_{P}/R_{*}$ & - & $0.0776_{-0.0048}^{+0.0046}$ \\
     ~~~ $a/R_{*}$ & - & $13.22_{-0.87}^{+0.97}$ \\
                           ~~~ $b$ & - & $0.43_{-0.28}^{+0.22}$ \\
                 ~~~ $T_{14}$ (days) & - & $0.0781_{-0.0089}^{+0.0086}$ \\
\sidehead{Transit 5:}
       ~~~ $T_C$ (\bjdtdb) & - & $6714.4316_{-0.0051}^{+0.0047}$ \\
~~~ $R_{P}/R_{*}$ & - & $0.0729_{-0.0053}^{+0.0050}$ \\
     ~~~ $a/R_{*}$ & - & $13.34_{-0.90}^{+0.99}$ \\
                           ~~~ $b$ & - & $0.42_{-0.28}^{+0.23}$ \\
                 ~~~ $T_{14}$ (days) & - & $0.0770_{-0.0096}^{+0.0088}$ \\
\sidehead{Transit 4:}
                ~~~ $T_C$ (\bjdtdb) & $6717.7641_{-0.0013}^{+0.0020}$ & $6707.76033_{-0.00058}^{+0.00061}$ \\
~~~ $R_{P}/R_{*}$ & $0.0825_{-0.0033}^{+0.0032}$ & $0.0744 \pm 0.0021$ \\
     ~~~ $a/R_{*}$ & $12.83_{-0.53}^{+0.78}$ & $13.64_{-0.67}^{+0.43}$ \\
                           ~~~  $b$ & $0.21_{-0.14}^{+0.17}$ & $0.24 \pm 0.16$ \\
                 ~~~ $T_{14}$ (days) & $0.0875_{-0.0059}^{+0.0032}$ & $0.0815_{-0.0014}^{+0.0015}$ \\
\sidehead{Transit 3:}
       ~~~ $T_C$ (\bjdtdb) & $6691.07394_{-0.00059}^{+0.00058}$ & $6684.40052_{-0.00081}^{+0.00075}$ \\
~~~ $R_{P}/R_{*}$ & $0.0820_{-0.0023}^{+0.0024}$ & $0.0780_{-0.0024}^{+0.0025}$ \\
     ~~~ $a/R_{*}$ & $12.68_{-0.84}^{+0.98}$ & $13.83_{-0.89}^{+0.62}$ \\
                           ~~~ $b$ & $0.551_{-0.12}^{+0.078}$ & $0.31_{-0.19}^{+0.16}$ \\
                 ~~~ $T_{14}$ (days) & $0.0782_{-0.0021}^{+0.0019}$ & $0.0791 \pm 0.0018$ \\
\sidehead{Transit 2:}
                    ~~~ $T_C$ (\bjdtdb) & $6681.0617_{-0.0020}^{+0.0019}$ & $6657.7023_{-0.0027}^{+0.0025}$ \\
~~~ $R_{P}/R_{*}$ & $0.0829_{-0.0053}^{+0.0052}$ & $0.0744_{-0.0035}^{+0.0032}$ \\
     ~~~ $a/R_{*}$ & $13.62_{-0.86}^{+0.85}$ & $12.79_{-0.73}^{+0.79}$ \\
                           ~~~ $b$ & $0.39_{-0.22}^{+0.17}$ & $0.18_{-0.13}^{+0.16}$ \\
                 ~~~ $T_{14}$ (days) & $0.0783_{-0.0043}^{+0.0041}$ & $0.0869_{-0.0052}^{+0.0057}$ \\
\sidehead{Transit 1 :}
       ~~~ $T_C$ (\bjdtdb) & $6661.0463_{-0.0017}^{+0.0016}$ & $6370.75698_{-0.00053}^{+0.00055}$ \\
~~~ $R_{P}/R_{*}$ & $0.0832_{-0.0057}^{+0.0060}$ & $0.0765_{-0.0030}^{+0.0027}$ \\
     ~~~ $a/R_{*}$ & $13.51_{-0.89}^{+1.0}$ & $13.69_{-0.86}^{+0.64}$ \\
                           ~~~ $b$  & $0.41_{-0.25}^{+0.18}$ & $0.33_{-0.19}^{+0.14}$ \\
                 ~~~ $T_{14}$ (days) & $0.076_{-0.010}^{+0.011}$ & $0.0799_{-0.0013}^{+0.0014}$ \\
           
\sidehead{Combined:}

     ~~~ P (days) &  $3.336525_{-0.000074}^{+0.000074}$ & $3.3366467 \pm 0.0000069$ \\

~~~ $R_{P}/R_{*}$ & $0.0833 \pm 0.0019$ & $0.0771_{-0.0011}^{+0.0012}$ \\
     ~~~ $a/R_{*}$ & $12.57_{-0.80}^{+0.95}$ & $13.47_{-0.82}^{+0.73}$ \\
                           ~~~ $b$ & $0.546_{-0.11}^{+0.074}$ & $0.36_{-0.19}^{+0.12}$ \\
                 ~~~ $T_{14}$ (days) & $0.0792_{-0.0017}^{+0.0017}$ & $0.08026_{-0.0010}^{+0.0012}$ \\

\enddata
\tablenotetext{-}{The transits in this table are listed in the same order, top to bottom, as in Figure \ref{fig:gpip}. The parameter symbols are defined as follows: $u_1$ = linear limb-darkening coefficient; $u_2$ = quadratic limb-darkening coefficient; $T_C$ = time of mid-transit - 2450000; $R_{P}/R_{*}$ = radius of planet in stellar radii; $a/R_{*}$ = semi-major axis in stellar radii; $b$ = impact parameter; $T_{14}$ = total transit duration; $P$ = orbital period.}
\label{tab:GJ3470b_LCOGT}
\end{deluxetable}

%%%%%%%%%%Kuiper Table%%%%%%%%%%

\begin{deluxetable}{lcc}
\tabletypesize{\footnotesize}
\tablewidth{9cm}
\tablecaption{Best-fit parameters found for GJ 3470\lowercase{b} Kuiper transits}
\tablehead{\colhead{~~~Parameter} & \colhead{Value (B)} & \colhead{Value (V)}}
\startdata

    ~~~ $\lambda$ (nm) & 437 & 545  \\
	~~~ ${\it u_1}$ & ${\it 0.421}\pm{\it 0.050}$ & ${\it 0.360}\pm{\it 0.050}$ \\
           ~~~ ${\it u_2}$ & ${\it 0.398}\pm{\it 0.050}$ & ${\it 0.411}\pm{\it 0.050}$ \\

       ~~~ $T_C$ (\bjdtdb)  & $6697.74734_{-0.00065}^{+0.00066}$ & $6677.72780_{-0.00058}^{+0.00054}$ \\
        ~~~ $R_{P}/R_{*}$ & $0.0827_{-0.0020}^{+0.0022}$ & $0.0770_{-0.0019}^{+0.0020}$ \\
     ~~~ $a/R_{*}$ & $13.60_{-0.90}^{+0.80}$& $13.42_{-0.98}^{+0.91}$ \\
                           ~~~ $b$ & $0.37_{-0.20}^{+0.13}$ & $0.41_{-0.20}^{+0.13}$ \\
                 ~~~ $T_{14}$ (days) & $0.0797_{-0.0017}^{+0.0018}$ & $0.0790_{-0.0014}^{+0.0015}$ \\

\enddata
%\tablenotetext{-}{}
\tablenotetext{-}{The parameter symbols are defined as follows: $u_1$ = linear limb-darkening coefficient; $u_2$ = quadratic limb-darkening coefficient; $T_C$ = time of mid-transit - 2450000; $R_{P}/R_{*}$ = radius of planet in stellar radii; $a/R_{*}$ = semi-major axis in stellar radii; $b$ = impact parameter; $T_{14}$ = total transit duration.}
\label{tab:GJ3470b_Kuiper}
\end{deluxetable}

%%%%%%%%%%Combined Table%%%%%%%%%%

\begin{deluxetable}{lc}
\tabletypesize{\footnotesize}
\tablewidth{9cm}
\tablecaption{Best-fit parameters found for all GJ 3470\lowercase{b} transits combined}
\tablehead{\colhead{~~~Parameter} & \colhead{Value }}
\startdata

       ~~~ $T_C$ (\bjdtdb)  & $6677.727712 \pm 0.00022 $  \\
        ~~~ $P$ & $3.3366413 \pm 0.0000060$ \\
     ~~~ $a/R_{*}$ & $12.92_{-0.65}^{+0.72}$ \\
                           ~~~ $b$ & $0.47_{-0.11}^{+0.074}$ \\
                 ~~~ $T_{14}$ (days) & $0.07992_{-0.00099}^{+0.00100}$  \\

\enddata
%\tablenotetext{-}{}
\tablenotetext{-}{The parameter symbols are defined as follows: $T_C$ = time of mid-transit - 2450000; $P$ = orbital period; $a/R_{*}$ = semi-major axis in stellar radii; $b$ = impact parameter; $T_{14}$ = total transit duration.}
\label{tab:GJ3470b_all}
\end{deluxetable}

%%%%%%%%%%Parameter tables%%%%%%%%%%
%%%%%%%%%%%%%%%%%%%%%%%%%%%%

\begin{figure*}[!t]
\begin{center}
\includegraphics[scale=0.28]{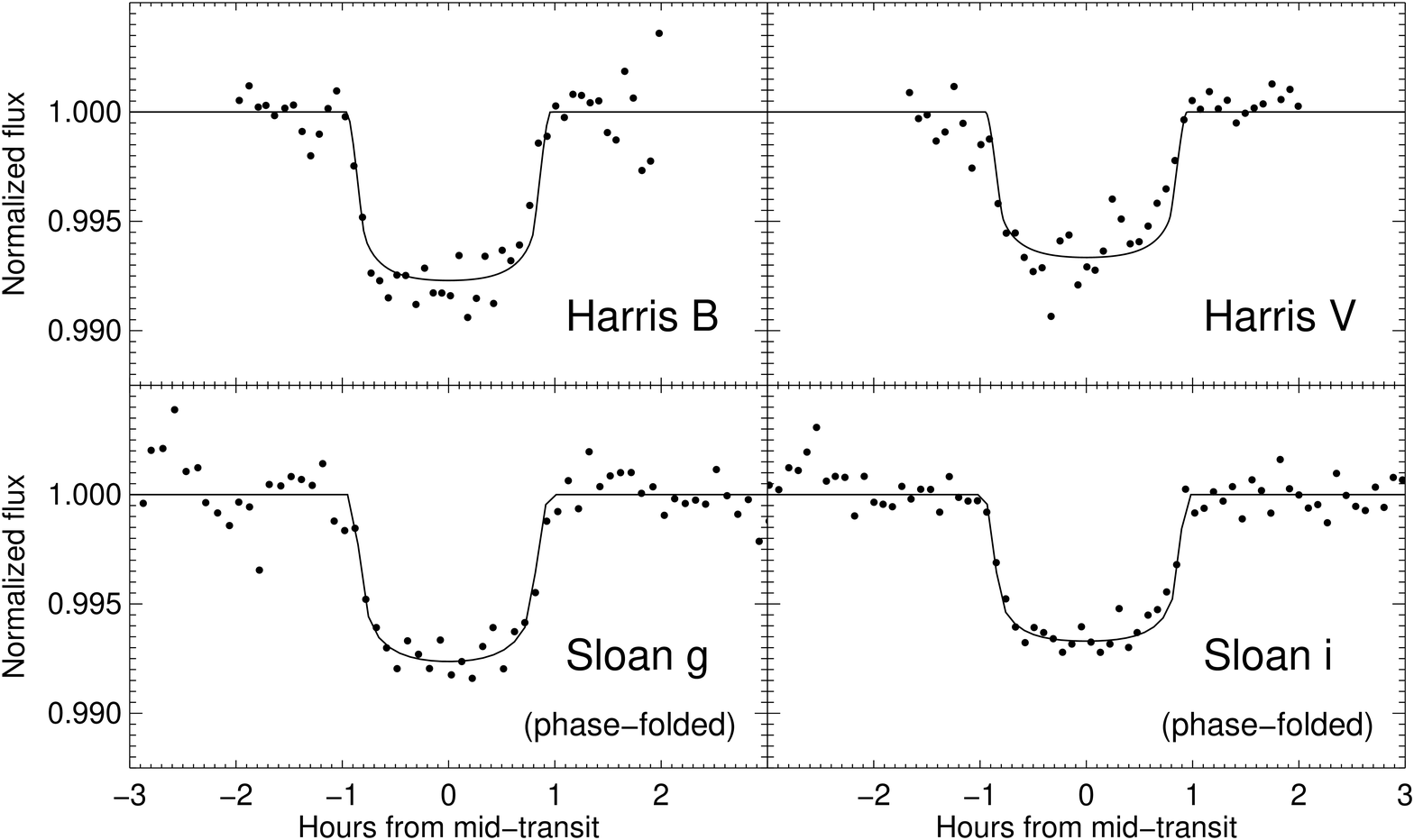}
\caption{{\it Upper:} Individual transits observed with the Kuiper telescope in Harris B on Feb. 8, 2014 ({\it left}), and in Harris B on January 14, 2014 ({\it right}), binned in 5-minute intervals. {\it Lower:} LCOGT Sloan g' ({\it left}) and Sloan i' ({\it right}) phase-folded transits, binned in phase in 5-minute intervals. Best-fitting transit models for each filter are shown. The associated best-fit parameters for the Kuiper and phase-folded LCOGT transits are listed in Tables \ref{tab:GJ3470b_Kuiper} and \ref{tab:GJ3470b_LCOGT}, respectively.}
\label{fig:alltrans}
\end{center}
\end{figure*}

We then fit all g' transits and all i' transits to obtain a set of transit parameters for each filter. Individual transit light curves were de-trended from different sets of variables, as described above. The mid-transit times were allowed to vary for each transit only to the extent to which they could still be fit with a linear ephemeris (i.e. transit timing variations were not allowed). All other parameters were not allowed to vary between transits in a given filter. The single B and V transits, and the phased g' and i' transits are plotted in Figure \ref{fig:alltrans}. Finally, we performed a global analysis of all twelve transits in order to obtain an updated ephemeris and precise values for $a/R_{*}$, $b$ and $T_{14}$, which can be found in Table \ref{tab:GJ3470b_all}. We note that all of our individual and global values of $P$, $a/R_{S}$, $b$ and $T_{14}$ agree with those of \cite{Dem13} to within 1$\sigma$, and with those of \cite{Bid14} to within 1 to 2$\sigma$. 

We also re-analysed the LBT light curves, fixing the limb darkening coefficients to theoretical values computed as described above, and scaling the photometric error bars to take into account the presence of correlated noise. Our reasons for this step are twofold: first, to provide an independent analysis of the discovery data and verify the statistical significance of the \cite{Nas13} detection; and second, to optimize the comparison between existing photometric observations in the visible wavelength regime by analyzing these observations in a consistent fashion. We indeed find a larger error bar for the planet-to-star radius ratio in the $U_{spec}$ transit light curve than reported by \cite{Nas13}. This leads to a 3.4$\sigma$ difference between the $U_{spec}$ and F97N20 $R_{P}/R_{S}$ values, significantly lower than the 6$\sigma$ difference determined by \cite{Nas13}. We note that the LBT observations, as well as those reported by \cite{Fuk13}, were also re-analyzed using a MCMC algorithm by \cite{Bid14}. We find that our $R_{P}/R_{S}$ measurements for the LBT light curves ($0.0823_{-0.0018}^{+0.0019}$ for $U_{spec}$ and $0.076_{-0.0004}^{+0.0004}$ for $F97N20$) differ slightly from those of \cite{Bid14}. Therefore, throughout this paper we use our own re-determination of the LBT $R_{P}/R_{S}$ values. 

%%%%%%%%%%%%%%%%%%%%%%%%%%%%%%%%%%%%%%%%%%%%%%%%%%%%%%%%%%%%%%%%%%%%

\section{Results}
\label{sec:Res}

Our analysis shows a 3$\sigma$ difference between the g' and i' $R_{P}/R_{S}$ values alone. When combined with our B- and V-band measurements as well as previously published values in the visible wavelength range ($\lambda <$ 1000 nm), the significance of the difference between planet-to-star radius ratio values below and above $\sim$650 nm increases even further. The hypothesis that this discrepancy has an astrophysical origin is strengthened by the fact that all of our g' $R_{P}/R_{S}$ values are consistently larger than those for the i' transits, as demonstrated in Figure \ref{fig:gpipcomp}. Incidentally, we do not see any systematic effects between transits observed at different LCOGT sites, indicating that data from the various LCOGT telescopes can be safely combined and analysed jointly.

One possible astrophysical cause for the difference in $R_p/R_s$ values could be stellar variability. We consider and discard this possibility for two reasons. First, \cite{Bid14} have shown that the low level of stellar variability is expected to introduce changes in the transit depth no greater than 5 $\times 10^{-5}$ for this planet (translating to variations of up to 3 $\times 10^{-4}$ in $R_{P}/R_{S}$). Second, our g' and i' transits acquired during the winter of 2013-2014 span nearly three cycles of the stellar rotation period (23.7 days; \citet{Bid14}) and are unevenly intermingled. This makes it very unlikely that the consistent difference in transit depths between the blue and red ends of the visible spectrum is due to stellar variability. We also consider unocculted spots as a potential cause for the observed increase in transit depth, as suggested by \cite{McC14}. Using the peak-to-peak amplitude in stellar variability of 0.01 as determined by \cite{Bid14} and \cite{Fuk13}, assuming the star spots are 300 K cooler than the stellar photosphere (3652 $\pm$ 50 K; \cite{Bid14}) and using equation 6 from \cite{McC14}, we find that the unocculted spots can only give rise to a difference in $R_p/R_s$ of no more than 0.0004. This is approximately 5$\%$ of the difference we observe (see Section \ref{sec:Disc} for the quantitative analysis). We conclude that the increase in $R_p/R_s$ is due to an increase in the radius of GJ 3470b toward shorter wavelengths.

\subsection{Atmospheric retrieval}

We use atmospheric retrieval to interpret our transmission spectrum. SCARLET, described in detail in \cite{Ben12} and \cite{Ben13}, starts by varying multiple parameters (metallicity, C/O ratio, cloud-top pressure, planetary radius at 1 bar and the Bond albedo) in a nested sampling framework. From these parameters, it generates molecular abundances in chemical equilibrium as well as temperature pressure (T-P) profiles. Molecular absorption is modelled through radiative transfer based on opacity look-up tables. Rayleigh scattering is included using the two-stream approximation. In this study, we included clouds as a grey opacity source that cuts off the transmission of starlight below the parameterized cloud-top pressure. For a given atmospheric composition, the Bond albedo is the dominant source of uncertainty in the (T-P) profile. The T-P profile determines the scale height, which in turn affects the mean molecular mass and the observed depths of the absorption features.

The nested algorithm computed several tens of thousands atmospheric models. It was initiated by randomly drawing 1000 active samples within the multidimensional parameter space. The active samples then iteratively converged toward regions of high likelihood. Convergence was obtained once the logarithm of the Bayesian evidence determined from the active sample changed by less than 0.0001. SCARLET is robust to multimodal posterior distributions and to degeneracies between parameters.

\begin{figure*}[!t]
\begin{center}
\includegraphics[scale=0.68]{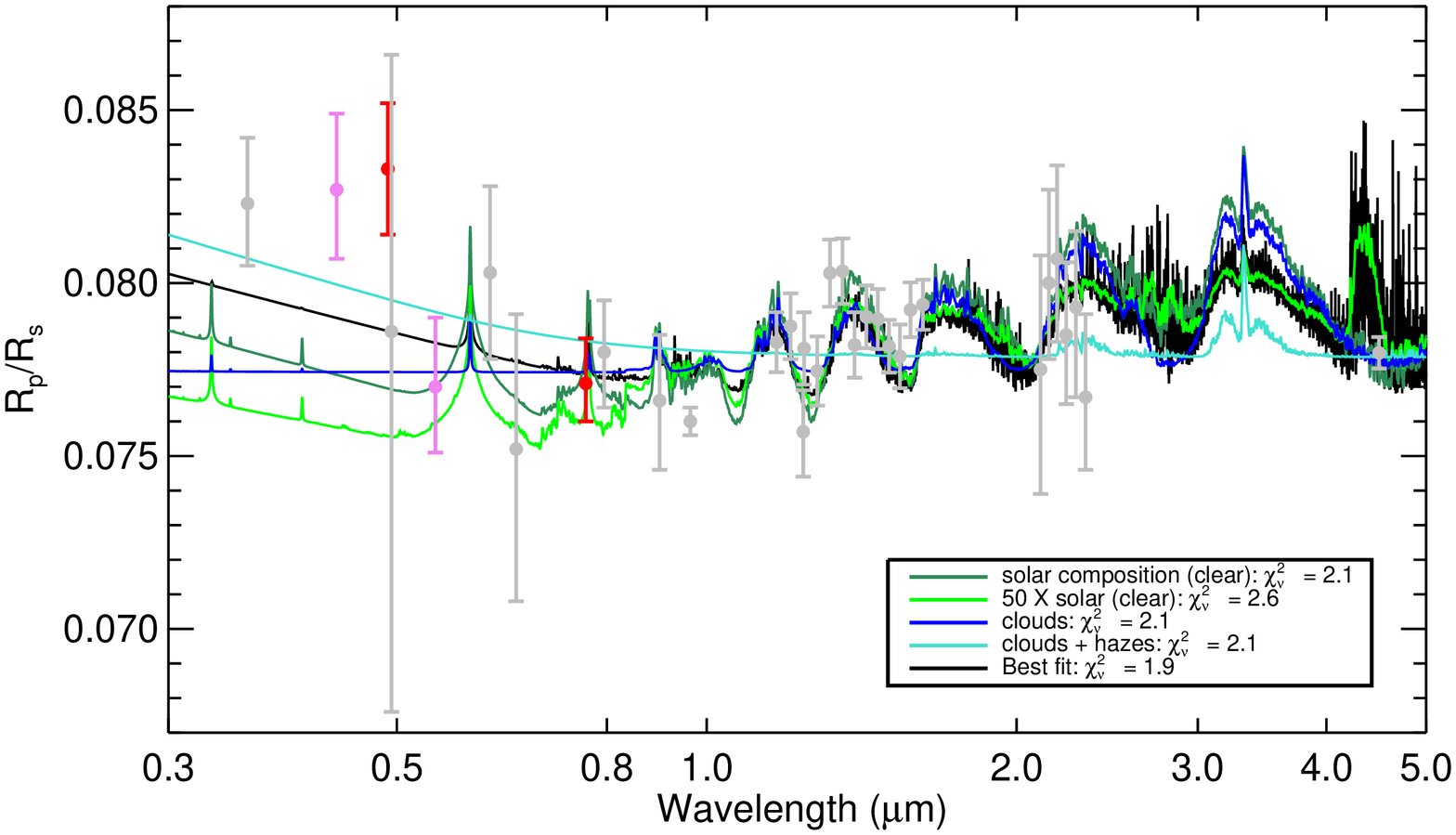}
\caption{Transmission spectrum for GJ 3470b. Red and magenta points are our LCOGT and Kuiper measurements, respectively. Gray points are previously published measurements taken from \cite{Dem13}, \cite{Bid14} (which includes a re-analysis of the data from \cite{Fuk13}), \cite{Ehr14} and \cite{Cro13}, as well as our re-analysis of the \cite{Nas13} LBT transits at 358 and 964 nm. The five colored lines show representative model spectra generated by the SCARLET algorithm. A colour version is available in the online version of the journal.}
\label{fig:allspec_model}
\end{center}
\end{figure*}

In Figure \ref{fig:allspec_model}, we show all published transmission spectra and broadband spectro-photometric transit measurements for GJ 3470b, the best-fit model spectrum as determined using the retrieval framework described above and four additional representative model spectra that provide similar fits. We also show reduced $\chi^2$ values for each model, using 31 degrees of freedom (32 data points, and a normalization parameter to match the average depth $R_p/R_s$ of our data).

\begin{figure}[!h]
\begin{center}
\includegraphics[scale=0.34]{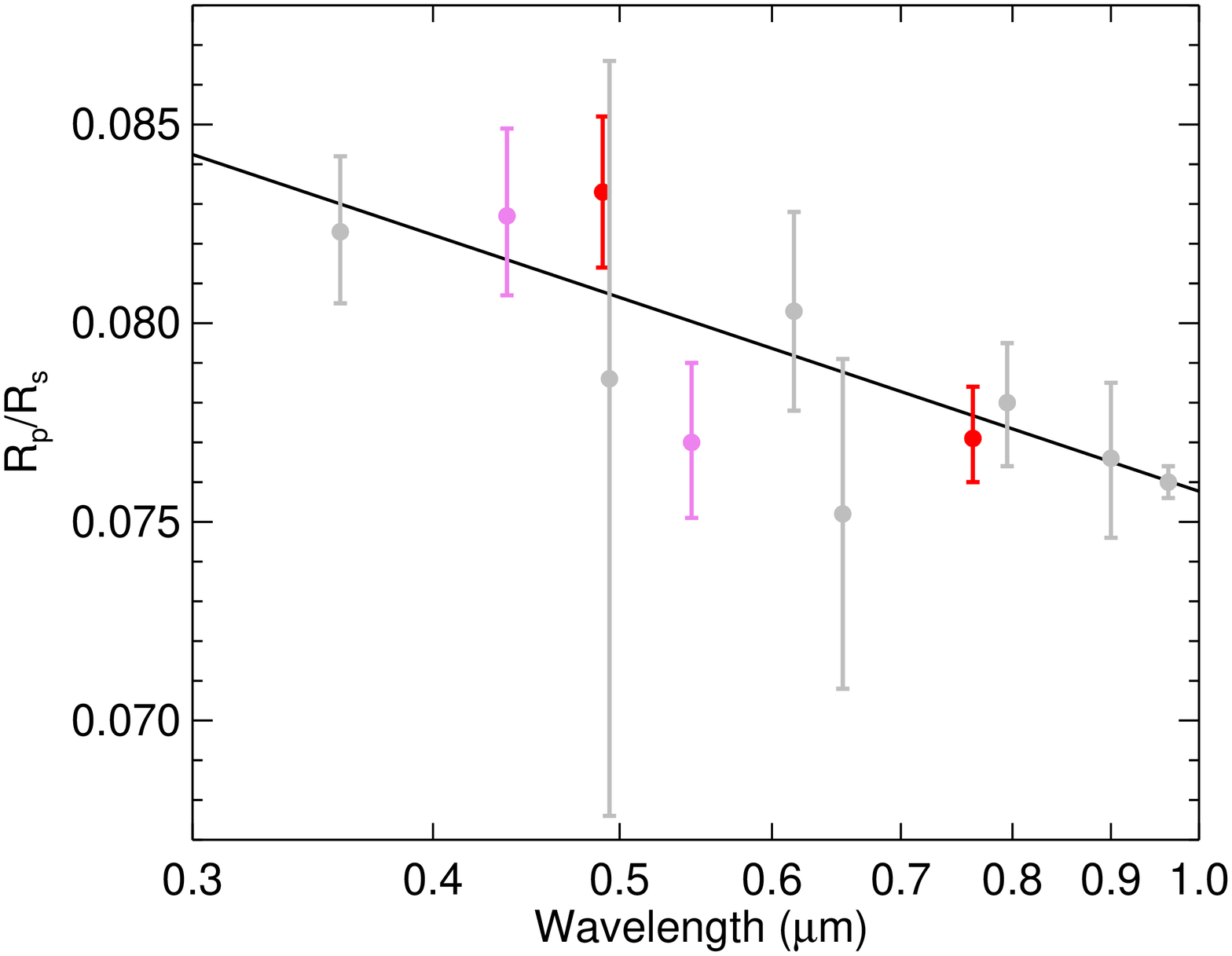}
\caption{Same as Figure \ref{fig:allspec_model}, but showing only the optical region. The solid line represents the best-fit to the data blueward of 1 $\mu$m. Its slope, combined with a value for $R_s$, determines the atmospheric scale height. A colour version is available in the online version of the journal.}
\label{fig:zoomspec_model}
\end{center}
\end{figure}

%%%%%%%%%%%%%%%%%%%%%%%%%%%%%%%%%%%%%%%%%%%%%%%%%%%%%%%%%%%%%%%%%%%%

\section{Discussion}
\label{sec:Disc}

Figure \ref{fig:allspec_model} suggests that all existing published spectroscopic data is reasonably consistent with a range of atmosphere models, of which we show a few representative ones. The best-fit model to the entire spectrum, in a $\chi^{2}$ sense, is a 50$\times$ solar metallicity one that includes water and methane absorption bands with some cloud and haze coverage. However, given the current data set, we favor a model that consists of a low mean molecular weight atmosphere covered by hazes and high altitude clouds (cyan line in Figure \ref{fig:allspec_model}). Our reasoning is as follows. First, we consider the fact that the WFC3 \citep{Ehr14} and MOSFIRE \citep{Cro13} observations spanning the 1 - 2.5 $\mu$m wavelength range prefer a flat spectrum to any that show physically motivated absorption features \citep{Ehr14}. Second, the steep Rayleigh slope in the visible regime is detected with high significance ($\chi^{2}_{\nu}$= 0.9 and 6.4, for the best-fit slope and a flat line, respectively; see below for description of fit). Clouds covering a H/He-dominated atmosphere would obscure molecular features in the near-infrared while hazes would give rise to a strong scattering feature in the visible. We note, as did \cite{Nas13}, that a clear solar composition atmosphere (dark green line in Figure \ref{fig:allspec_model}) alone is not sufficient to reproduce the observed steep slope while also providing a reasonable fit to the data points redward of 1 $\mu$m.

The main characteristic of the transmission spectrum that prevents a ``clouds and hazes'' model from providing the best fit to the data is an offset between the median of the near- and mid-IR data points, and the level where the Rayleigh scattering slope appears to flatten. We consider two factors that may cause this discrepancy. One is stellar variability due to spots, which may induce a scatter of up to 0.008 in the $R_{p}/R_{s}$ values, particularly those in the visible regime (the contrast between star spots and the stellar photosphere decreases toward longer wavelengths). However, stellar variability alone cannot explain this difference. The offset may also arise from differences between the reduction and analysis procedures carried out for each of the low-resolution spectra and the broadband spectrophotometric measurements by the groups who acquired the data. As an illustration, \cite{Bid14} re-analyzed the previously published transits of \cite{Fuk13} and \cite{Nas13} and obtained values that differed from the original values by up to 2$\sigma$, as well as different uncertainties on the transit depths. We note that \cite{Nik13} find an offset between the HST STIS and WFC3 HAT-P-1b transmission spectra they analysed, though in their case the level of the spectra longward of 1.0 $\mu$m (WFC3) is {\it lower} than that of the spectra shortward of 1.0 $\mu$m (STIS). They have also found that stellar variability could not completely explain the difference between the mean levels of the spectra, but unknown instrumental systematics remain a possibility. 

Following the procedure of \cite{Nas13}, we have also constrained the mean molecular weight ($\mu$) of the planet's atmosphere using the following equation:

\begin{eqnarray}
H = \frac{1}{\alpha}\frac{\mathrm{d}R_{p}}{\mathrm{dln}\lambda} = \frac{kT_{eq}}{\mu g},
\end{eqnarray}

\noindent where $H$ is the atmospheric scale height, $k$ is Boltzman's constant and $\alpha = -4$ for Rayleigh scattering. We measured the slope of the observed Rayleigh scattering feature ($dR_p/R_s\mathrm{dln}\lambda$) through a weighted linear least squares fit, obtaining a best-fit value of

\begin{eqnarray}
\frac{\mathrm{d}R_{p}}{R_s\mathrm{dln}\lambda} = -0.0070 \pm 0.0013.
\end{eqnarray}

\noindent Figure \ref{fig:zoomspec_model} shows the visible region of the transmission spectrum and the best-fit slope. We then used this value together with the stellar radius ($R_{s} = 0.48  \pm 0.04 R_{\odot}$) and the planet's gravity ($g$ = 676.1 $\pm$ 171 cm s$^{-2}$) from \cite{Bid14}, as well as its equilibrium temperature ($T_{eq} = (1-A_B)^{1/4}$ 691.6 $\pm$ 15 K) estimated from stellar and transit parameters in the same publication, to derive $\mu$ for two limiting cases of the Bond albedo ($A_B$): 0 and 0.3. We obtain a mean molecular weight of 1.47 $\pm$ 0.48 for $A_B$=0 and 1.35 $\pm$ 0.44 for $A_B$=0.3 \footnote{Theoretical models \citep{Spi10} and {\it Kepler} observations \citep{Est13} suggest that 0.3 is a reasonable upper limit, in general, on the Bond albedo of hot gaseous exoplanets.}. Our values agree with those found by \cite{Nas13} and are consistent with a H/H$_2$/He-dominated atmosphere, confirming our conclusions from the beginning of this Section. 

Nevertheless, the existing data set does not constrain the atmospheric composition of GJ~3470b beyond indicating a low-metallicity atmosphere. One way forward is through additional transit spectroscopy observations in the near- and mid-infrared, where the absorption features expected from water and carbon molecules may be detected as long as any clouds are at sufficiently low altitude. If with more precise measurements the spectrum remains featureless in this wavelength range, this would confirm the presence of clouds high in the atmosphere, but would also limit the chances of constraining this planet's atmospheric composition any further. 

\section{Conclusion}
\label{sec:Conc}

We have observed several transits of GJ 3470b, a warm Neptune analog around an early M dwarf, in four different bands with the LCOGT and Kuiper telescopes. Our analysis of the resulting photometry shows a marked increase in the planetary radius from the red toward the blue ends of the visible wavelength range. Our LCOGT g' and i' band measurements are based on 4 and 6 transits, respectively. These transits were observed at four different sites and are unevenly spaced over a period spanning over three stellar rotation cycles. The g' transits consistently have a larger $R_p/R_s$ value than the i' transits, allowing us to rule out stellar activity and site-specific systematics as the source of this variation.

We conclude that the rise toward the blue end of the transmission spectrum is due to an increase in the planetary radius, indicative of Rayleigh scattering in the planet's atmosphere. We find that the most plausible scenario is a H/He-dominated atmosphere covered by high-altitude clouds and hazes, which obscure absorption features in the near-IR but give rise to a steep Rayleigh scattering slope in the visible, respectively. With the visible slope of GJ 3470b's transmission spectrum confirmed by our LCOGT and Kuiper measurements, the next step consists of acquiring more infrared observations in order to verify whether the spectrum is truly featureless redward of 1 $\mu$m.

Our result is the first high-confidence detection of an exoplanet atmospheric feature using observations taken only with 1.0 to 2.0m telescopes. This work demonstrates the importance of the role that ground-based, meter-class telescopes can play in the characterization of exoplanet atmospheres. A search for Rayleigh scattering in the atmospheres of the most promising, low-density exoplanets with deep transits becomes conceivable. Such measurements would be particularly valuable for those planets whose transmission spectrum at longer wavelengths does not show any sign of absorption features, but even when that is not the case, they can provide an independent determination of the atmospheric scale height and composition. Meaningful constraints on these parameters could be obtained with only a few transits observed in each of two filters widely separated in wavelength (i.e. Sloan g' and i', Bessell B and I, etc.), depending on the planet-to-star radius ratio and the planetary density. However, we strongly recommend that such observations be taken in the same sets of (preferably standard) filters, in order to facilitate comparison between independent sets of measurements and the eventual confirmation of a tentative signal.

%%%%%%%%%%%%%%%%%%%%%%%%%%%%%%%%%%%%%%%%%%%%%%%%%%%%%%%%%%%%%%%%%%%%

\section*{Acknowledgements}

The authors thank David Ehrenreich, Heather Knutson, Jacob Bean and Rob Siverd for fruitful discussions. We are grateful to the anonymous referee for suggestions which have led to improvements in the paper. This research makes use of observations from the LCOGT network, and of the SIMBAD database, operated at CDS, Strasbourg, France. I. J. M. C.'s work was performed under a contract with the California Institute of Technology funded by NASA through the Sagan Fellowship Program.

%\clearpage

%%%%%%%%%%%%%%%%%%%%%%%%%%%%%%%%%%%%%%%%%%%%%%%%%%

\bibliographystyle{apj}
\bibliography{GJ3470_ms}

\end{document}